\documentclass{acm_proc_article-sp}

\usepackage{hyperref}
\begin{document}

\title{Swift: task-based hydrodynamics and gravity for cosmological simulations}

\newcommand{\swift}{{\sc swift}}
\newcommand{\quicksched}{{\sc QuickSched}}
\numberofauthors{4}

\author{
\alignauthor
Tom Theuns\\
       \affaddr{Institute for Computational Cosmology}\\
       \affaddr{Department of Physics}\\
       \affaddr{Durham University}\\
       \affaddr{Durham DH1 3LE, UK}      
\alignauthor
Aidan Chalk\\
       \affaddr{School of Engineering and Computing Sciences}\\
       \affaddr{Durham University}\\
       \affaddr{Durham DH1 3LE, UK}
\alignauthor Matthieu Schaller\\
       \affaddr{Institute for Computational Cosmology}\\
       \affaddr{Department of Physics}\\
       \affaddr{Durham University}\\
       \affaddr{Durham DH1 3LE, UK}
\and
\alignauthor Pedro Gonnet\\
       \affaddr{School of Engineering and Computing Sciences}\\
       \affaddr{Durham University}\\
       \affaddr{Durham DH1 3LE, UK}\\
       \affaddr{and}\\
       \affaddr{Google Switzerland GmbH}\\
       \affaddr{Brandschenkestr. 110}\\
       \affaddr{8002 Zurich, Switzerland}
        }

\date{29 July 2015}

\maketitle
\begin{abstract}
Simulations of galaxy formation follow the gravitational and
hydrodynamical interactions between gas, stars and dark matter through
cosmic time. The huge dynamic range of such calculations severely
limits strong scaling behaviour of the community codes in use, with
load-imbalance, cache inefficiencies and poor vectorisation limiting
performance. The new \swift\ code exploits task-based parallelism
designed for many-core compute nodes interacting via MPI using
asynchronous communication to improve speed and scaling. A graph-based
domain decomposition schedules interdependent tasks over available
resources. Strong scaling tests on realistic particle distributions
yield excellent parallel efficiency, and efficient cache usage
provides a large speedup compared to current codes even on a single
core. {\sc swift} is designed to be easy to use by shielding the
astronomer from computational details such as the construction of the
tasks or MPI communication. The techniques and algorithms used in {\sc
  swift} may benefit other computational physics areas as well, for
example that of compressible hydrodynamics. For details of this
open-source project, see \url{www.swiftsim.com}
\end{abstract}



\section{Introduction}
The main aim of cosmological simulations of the formation of
structures in the Universe is to understand which physical processes
play in role in how galaxies form and evolve. For example, what
determines whether a galaxy becomes a spiral or an elliptical? What is
the origin of the morphology-density relation - the observation that
elliptical galaxies cluster much more strongly than spirals? What sets
the colours of galaxies? How does the rate of galaxy formation evolve
over cosmic time? What is the nature of high-redshift galaxies? A
better understanding of these processes will be required to take full
advantage of the rich data sets being collected now, or promised by
future observatories such as the James Webb space
telescope\footnote{\url{http://www.jwst.nasa.gov/}}, ESO's
Extremely-Large
telescope \footnote{\url{http://www.eso.org/public/teles-instr/e-elt/}}
or the Square Kilometre
Array \footnote{\url{https://www.skatelescope.org/}}.

Such cosmological simulations start from initial conditions motivated
by observations of the cosmic microwave background (CMB). The CMB
provides a directly observable imprint of the small density
fluctuations that will eventually grow due to gravity into galaxies
and clusters of galaxies today. In an expanding universe, regions
which are slightly over-dense become denser and eventually collapse
due to the self-gravity of their dark matter. These collapsing \lq
halos\rq\ accrete gas that cools radiatively and makes stars. The
simulations follow the build-up of the dark matter halos and the
accretion, shock-heating, and radiative cooling of the gas onto halos.

The gas densities above which stars form are orders of magnitude
higher than the typical density in a galaxy and this large dynamic
range is one of the most challenging aspects of these
computations. The radiation and winds of recently formed stars, and
the energy injected by super nova explosions, strongly limit the rate
at which a galaxy's gas is turned into stars. As a result, only $\sim
17$ per cent of all gas in the Universe has been converted into stars
to date \cite{Fukugita98}. The tremendous dynamic range in mass,
length and time, between gas accreting onto a halo and turning into
stars prevents simulations to model these crucial processes in
detail. \lq Subgrid\rq\ schemes are therefore used to model processes
that cannot (yet) be resolved numerically, not unlike what is done in
other multi-scale calculations such as for example weather or climate
modelling. Limiting the impact of these subgrid models by actually
resolving some of the underlying physics is a tremendously exciting
and computationally demanding challenge for the exascale era.

Current cosmological simulations often take months to run on hundreds
to many thousands of cores. For example the recent {\sc EAGLE}
simulation \cite{Schaye15} took 45 days to run on 4000 cores of the
Durham Data Centric Cluster, part of the {\sc DIRAC}
infrastructure\footnote{\url{http://www.stfc.ac.uk/1263.aspx}}, and
the simulation suite used nearly 40M core hours on the {\sc curie}
machine using a {\sc prace}\footnote{\url{http://www.prace-ri.eu/}}
allocation of computer time. Such long run times are currently
limiting scientific progress.

This paper discusses the {\sc swift} code that is designed to
over-come some of the limitations of community codes widely used in
cosmology, in particular improving load-balance, cache-usage, and
vectorisation. It also intends to shield the astronomer who intends to
implement and test subgrid schemes from the underlying computational
details.

\section{Cosmological gas dynamics}
This section provides a brief overview of the equations being
integrated. Calculations are performed in co-moving coordinates ${\bf
  x}$ say for position, related to physical coordinates ${\bf r}$ by
the time-dependent scale factor $a(t)$, ${\bf r}=a{\bf x}$ (see for
example \cite{Peebles93}), but we will ignoring these details
here. Performing these calculations using a Lagrangian scheme where
the fluid is represented by a set of particles that move with the
fluid's speed is very advantageous, because the flow speeds are very
large due to the large (gravitational) motions of forming galaxies.

Smoothed particle hydrodynamics (SPH, \cite{Gingold83, Lucy77}) is
such a Langrangian scheme in which values for fluid variables are
interpolated from a disordered particle distribution using kernel
interpolation. For example the density $\rho$ and pressure gradient
$\nabla p$ at the location ${\bf r}_i$ of particle $i$ are computed
with equations of the form
\begin{eqnarray}
\label{eq:rho}
\rho({\bf r}_i) &= & \sum_j m_j\,W({|{\bf r_i}-{\bf r_j}|\over h_i})\,,\\
\label{eq:pres}
\nabla p ({\bf r}_i) &= & \sum m_j \left(
      {p({\bf r}_i)\over \rho({\bf r}_i)^2} + {p({\bf r}_j)\over \rho({\bf r}_j)^2}
      \right)
      \nabla W({|{\bf r_i}-{\bf r_j}|\over h_i})
\end{eqnarray}
where $m_j$ is the mass of particle $j$ and $W$ is a bell-shaped
kernel with compact support, $W(q)=0$ for $q>1$. The smoothing length
$h_i$ is computed such that a given weighted number of particles
contributes to the sum. The pressure is found from the density and
temperature using an equation of state. Note that we need to evaluate
the density for each particle before we can compute the pressure
gradient. Several variations of Eq.~(\ref{eq:pres}) exist, we use this
particular form here to illustrate the type of sums to be computed,
{\sc swift} implements the more accurate version used in {\sc gadget
  2} \cite{Springel05}.

Gravitational accelerations are calculated as,
\begin{equation}
\label{eq:grav}
{\bf a}_i =-{\rm G}\sum_{j\ne i}\,{m_j\over |{\bf r_i}-{\bf r_j}|^3}\,({\bf r}_i-{\bf r}_j)\,,
\end{equation}
with extra terms (not discussed here) to represent periodic images
such that the simulated volume is periodically replicated (the Ewald
summation familiar from solid state physics).

Given the initial state of the system, specified by position and
velocities of all particles, particles are marched forward in time
using velocities to update positions and accelerations to update
velocities.  Most of the calculation time is spent in evaluating the
hydrodynamical and gravitational forces. The popular {\sc gadget}
\cite{Springel05} and {\sc gasoline} \cite{Wadsley04} codes use a tree
to find neighbours for evaluating the sum in
Eqs.~(\ref{eq:rho}-\ref{eq:pres}). These codes split the gravitational
force from Eq.~(\ref{eq:grav}) into a contribution from nearby
particles evaluated using a tree following \cite{Barnes86}, and
contribution from distant particles evaluated using a mesh, as in the
P3M scheme described in detail in \cite{Hockney88}, see
\cite{Efstathiou85} for the application in cosmology. The particles
are distributed over the computational volume using a space-filling
curve to attempt to preserve locality which reduces MPI communication
needed if neighbour particles are not held on the same MPI task. Such
\lq domain decomposition\rq\ also takes significant compute time. How
this issues are handled in {\sc swift} is described next.

\section{Task-based calculations}
\subsection{SPH}
\begin{figure}
\centering
\includegraphics[width=\columnwidth]{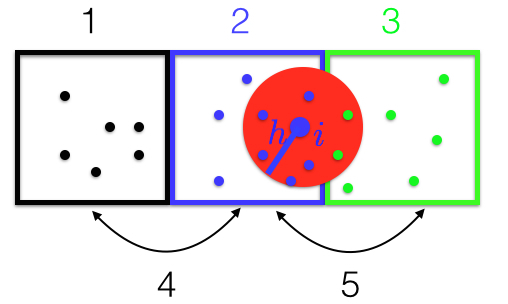}
\caption{Illustration of neighbour finding on a mesh. Five tasks are indicated, numbers 1-3 compute densities from pairs of particles in cells 1-3, whereas tasks 4 and 5 compute densities between particles pairs in neighbouring cells.}
\label{fig:sph}
\end{figure}
\begin{figure}
\centering
\includegraphics[width=\columnwidth]{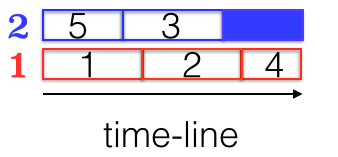}
\caption{Execution of the 5 tasks (labelled 1-5) illustrated in Fig.\ref{fig:sph}, by two threads (labelled 1 and 2 and coloured red and blue, respectively) with conflicts. Thread 1 starts executing task 1, while thread 2 executes task 5, locking tasks 2, 3 and 4. When thread 2 completes task 5, it immediately starts executing task 3. Thread 1 can execute task 2 locking task 4 when task 1 is completed. However thread 2 cannot start executing task 4 as long as task 2 is not completed, since tasks 2 and 4 conflict.}
\label{fig:task-graph}
\end{figure}
\begin{figure}
\centering
\includegraphics[width=\columnwidth]{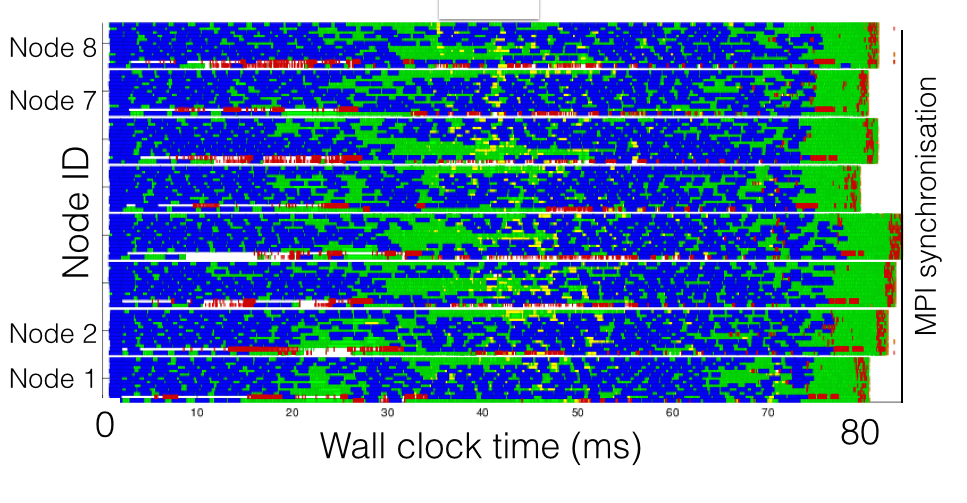}
\caption{Task time-line for {\sc swift} SPH calculation, running on 8 nodes (thick bands) with 12 cores (thin bands) each. Different colours corresponds to different tasks, for example red refers to communication. As the calculation progresses, each core is executing tasks mostly independent of other cores, with little idle time lost due to MPI synchronisation at the end of the time step.}
\label{fig:taskline}
\end{figure}

{\sc Swift} identifies potential neighbours by organising particles in
cubic cells as illustrated in Fig.\ref{fig:sph} (drawn in 1 dimension
for simplicity). By choosing the cell size of the mesh to be larger
than the smoothing length $h$ of all particles in that cell guarantees
that particles within $h_i$ of the fat blue particle in the figure can
be found either in the same cell (blue, labelled \lq 2\rq), or in one
of the two neighbouring cells (black and green, labelled \lq 1\rq\ and
\lq 3\rq\ respectively). Given the large dynamic range in $h$, such a
mesh needs to be adaptive. The density calculation of
Eq.~(\ref{eq:rho}) for particle $i$ now involves three steps: find
neighbours of $i$ in each of the three cells (in the figure, these are
particles within the red circle with radius $h_i$).

In {\sc Swift}, each of these calculations is executed by separate
{\bf tasks}. In the simple case illustrated in Fig.\ref{fig:sph} there
are two types: tasks that involve evaluating Eq.~(\ref{eq:rho}) for
pairs of particles in the {\em same} cell (labelled 1-3), and tasks
that involve evaluating Eq.~(\ref{eq:rho}) for pairs of particles in
neighbouring cells (labelled 4 and 5). To avoid race conditions, some
tasks cannot be performed simultaneously, in this particular case
tasks 4 and 5 conflict will each other, 4 conflicts with 1 and 2, and
5 with 2 and 3 . The task scheduling in {\sc swift} therefore should
be able to handle both conflicts and dependencies.

How these 5 tasks could be executed by two threads is illustrated in
Fig.~\ref{fig:task-graph}. At the start, threads pick tasks
independently, locking those tasks that conflict with them. In this
example, thread 1 executes task 1, and thread 2 executes task 5
(locking tasks 2 and 3). When thread 2 completes task 5 it unlocks
tasks 2 and 3, and starts executing task 3 (locking task 4). When task
3 is finished, thread 2 is idle because the remaining task 4 conflicts
with task 2 being executed by thread 1. One of the threads (in the
illustration thread 1) finishes off the work.

The efficiency of the tasks themselves can be improved by {\em
  sorting} \cite{Gonnet13,Gonnet14}. Indeed, consider again the fat
blue particle $i$ in Fig.\ref{fig:sph} when task 5 is executed. If we
were to sweep through the green particles in cell 3 {\em from left to
  right}, we would find that the fourth green particle no longer
contributes to the density since it is outside the red circle. There
is therefore no reason to even check if any of the other green
particles is inside the red circle, since these are even further away
from particle $i$ in the horizontal direction.

The {\sc swift} SPH implementation contains several similar \lq
kernels\rq\ that calculate the interaction between two particles (for
example individual terms in Eq.~(\ref{eq:rho}) or in
Eq~(\ref{eq:pres})). Exposing these basic routine to the user greatly
simplifies adapting the code to the user's wishes, for example in
making changes to the basic SPH algorithm. This kernel is called for a
range of particles that are in the same cell. Cache-misses are
minimised by making sure these particles are nearly contiguous in
memory. Bunching particles in cells is then also advantages for
vectorising, either using intrinsics, or by using pragma's that allow
the compiler to known that these calculations can be vectorised.

With sorting tasks, density tasks, and pressure gradients tasks (and
gravity tasks, described next) combined for all cells, a science run
will typically contain hundreds or even millions of tasks. Individual
threads on a many-core node can thus all be executing tasks as long as
these do not conflict with each other, using task stealing to grab a
new task as soon as their current task is completed. Once a thread
grabs a new task, it blocks those tasks that conflict with it. In
addition to conflicts, the {\sc swift} task engine also handles task
{\em dependencies} - for example the density of particles in a cell
and its neighbouring cell should have been computed before pressure
gradients can be computed.

Running this task-based parallelisation across MPI tasks introduces
relatively minimal additional complexity. If neighbouring cells are
assigned to different MPI tasks, {\sc swift} will generate extra
communication tasks that exchange the contents of individual cells
using asynchronous communication. The distribution of particles (or
rather cells) across MPI tasks is based on the total costs of tasks -
assigning similar work to each MPI task - while aiming to minimise
communication that results from spatially non-contiguous particle
distributions. Generating and scheduling the inter-dependent tasks is
performed in a similar way as is done in the {\sc QuickSched} library
\cite{Gonnet14} using the {\sc metis} library \cite{Karypis98} to
partition tasks over MPI tasks. An example is shown in
Fig.\ref{fig:taskline} (a realistic version of
Fig.~\ref{fig:task-graph}), which shows a time-line of how 8 nodes of
12 threads each execute a set of tasks using MPI across nodes. Running
on a realistic particle distribution, {\sc swift} achieves 60 per cent
parallel efficiency in a strong scaling test increasing the core count
from one to 1024 (see Fig.~\ref{fig:sphscaling}, see also
\cite{Gonnet14}).

Using cells to organise particles spatially and identify potential
neighbours may at first sight seem very different from using a tree as
in the {\sc Gadget 2} or {\sc Gasoline} codes. However the algorithms
are actually surprisingly similar once one limits the depth of the
tree to cells that contain $\sim 100$ particles as is the case in {\sc
  swift}. How to find neighbouring {\em cells} in {\sc swift} is
actually also performed using a tree.

\subsection{Gravity}
\begin{figure}
\centering
\includegraphics[width=0.8\columnwidth]{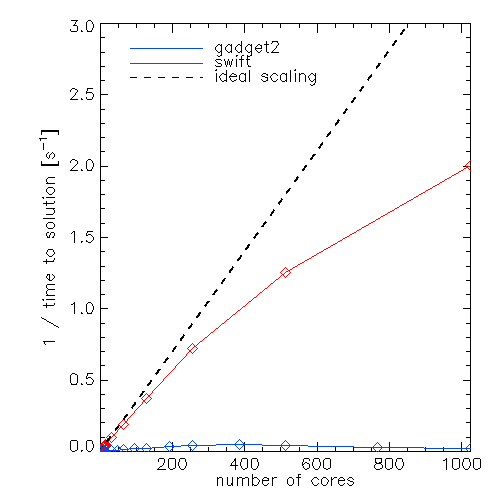}
\caption{Time to solution strong scaling test of the SPH implementation in {\sc swift} compared to {\sc Gadget 2} for a realistic particle distribution with 51 million particles taken from a cosmological volume. Scaling is shown from 1 to 1024 cores (64 nodes with 16 cores each). {\sc swift} uses 16 threads per core, {\sc gadget 2} uses MPI also within a node. {\sc swift} reaches 60 per cent parallel efficiency for strong scaling from 1 to 1024 cores.}
\label{fig:sphscaling}
\end{figure}

\begin{figure}
\centering
\includegraphics[width=0.8\columnwidth]{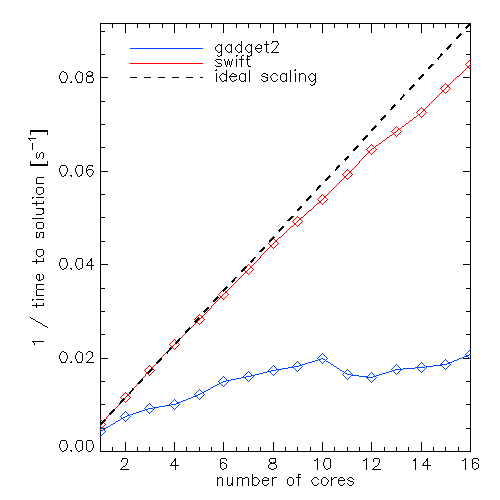}
\caption{Time to solution strong scaling test of the Barnes-Hut gravity implementation in {\sc swift} compared to {\sc Gadget 2} for a 10M highly-clustered particle distribution on a single node. Increasing the thread count from 1 to 16 reduces the time to solution in {\sc swift} by a factor 14, a 90 per cent efficiency (red line). Increasing the number of MPI-tasks for {\sc gadget-2} from 1 to 16 decrease the time to solution by factor of 5.}
\label{fig:gravscaling}
\end{figure}
Currently {\sc swift} implements the Barnes-Hut tree code algorithm
\cite{Barnes86} for evaluating the gravitational acceleration from
Eq.~(\ref{eq:grav}), with some modifications described below. The
Barnes-Hut algorithm divides the simulation volume spatially and
recursively in smaller cells. Such a division is very well suited for
evaluating gravitational interactions. Indeed consider a particle $i$
at some distance from a tree node. A good approximation for the
contribution of that node to ${\bf a}_i$ can be obtained using a
multipole expansion, for example representing all the particles in the
node by their monopole, as long as the distance particle-node is large
compared to the extent of the node. If the distance is small, the node
is split in its daughter cells, and the algorithm recurs.
	
This Barnes-Hut algorithm decreases the computational cost of
evaluating ${\bf a}_i$ for all particles from order $N^2$ to order
$N\log(N)$ \cite{Barnes86}. Note that two particles that are spatially
close are likely to execute nearly identical tree walks. In practise
most of the compute time is now spent in the tree walk (rather than
evaluating actual accelerations).
		
We implemented three optimisations of this algorithm in {\sc
  swift}. Firstly we limit the depth of the tree from leaf nodes that
contain a single particle (as in {\sc gadget}) to cells with $\sim
100$ particles. This is because the tree walk is not very efficient
for {\em small} numbers of particles.
	
Secondly we do not start a tree walk for each particle from the root
node, but rather walk the tree walk for {\em nodes}. For each set of
nodes, we decide whether they are sufficiently distant to compute
forces using multipoles, or they should be split in their daughter
nodes recursively. Doing so results in a list of tasks, those in which
particles in one node interact with the multipole of another node, or
those where all particles in one node interact with all particles in a
nearby node. The latter task is implemented efficiently using the same
task-based approach as used for SPH in the previous section.
	
Thirdly we use quadrupoles rather than monopoles. This increases time
to solution minimally yet make the accelerations more accurate.

The speed and scaling of the tree implementation in {\sc swift} is
compared to that of {\sc gadget 2} in Fig.~\ref{fig:gravscaling}, in
which ${\bf a}_i$ is calculated for each of 10M particles taken from
the same snapshot of an {\sc eagle} simulation as used in
Fig.~\ref{fig:sphscaling} (a very clustered distribution of
particles). The speed of {\sc swift} is close to that of {\sc gadget
  2} when run on a single core, and the scaling up to 16 threads is
close to ideal (parallel efficiency of 90 per cent).  The public
version of {\sc gadget 2} does not have multi-threading, and the
scaling shown is when increasing the number of MPI tasks using one
core per task.

\section{Conclusions}
We have implemented smoothed particle hydrodynamics (SPH) and a
Barnes-Hut tree-code for self-gravity in the cosmological
hydrodynamics code {\sc swift}. By grouping nearby particles in cells,
the calculation is broken-up into very many short and inter-dependent
tasks, whereby a single task processes particles within a cell, or
between pairs of cells.  Task dependencies and conflicts are encoded
in the application.  Using cells improves cache efficiency and
simplifies vectorisation. The tasks are distributed across nodes, with
individual threads using task-stealing within a node, and
communication being performed asynchronously between nodes. We find
that such task-based parallelism is well suited to take advantage of
the multiple levels of parallelism of modern many-core super
computers. Applied to a realistic particle distribution, {\sc swift}'s
SPH implementation reaches a parallel efficiency of 60 per cent in a
strong scaling test when increasing core count from 1 to 1024, and
better than 90 per cent on a single 16-core node for
gravity. Individual physics routines, for example those that evaluate
interactions between two particles, are implemented in simple kernels
to shield the physicist from the intricacies of tasks or MPI
communications. {\sc swift} is an open-source project,
\url{www.swiftsim.com}.

\section*{Acknowledgments}
This work used the DiRAC Data Centric system at Durham University,
operated by the Institute for Computational Cosmology on behalf of the
STFC DiRAC HPC Facility (\url{www.dirac.ac.uk}). This equipment was
funded by BIS National E-infrastructure capital grant ST/K00042X/1,
STFC capital grant ST/H008519/1, and STFC DiRAC Operations grant
ST/K003267/1 and Durham University. DiRAC is part of the National
E-Infrastructure. This work was supported by the Science and
Technology Facilities Council [ST/F001166/1] and the European Research
Council under the European Union's ERC Grant agreements 267291
Cosmiway, by the Interuniversity Attraction Poles Programme initiated
by the Belgian Science Policy ([AP P7/08 CHARM]), and by {\sc intel}
through establishment of the ICC as an {\sc intel} parallel computing
centre (IPCC).

\bibliographystyle{abbrv}
\bibliography{paper}  
\end{document}